# Enhancement of photovoltaic efficiency by insertion of a polyoxometalate layer at the anode of an organic solar cell†

M. Alaaeddine,[a,b] Q. Zhu,[a,b] D. Fichou,[a,c] G. Izzet,[a,c] J. E. Rault,[d] N. Barrett,[e] A. Proust*[a] and L. Tortech*[a,b]

In this article the Wells-Dawson polyoxometalate $K_6[P_2W_{18}O_{62}]$ is grown as an interfacial layer between indium tin oxide and bulk heterojunction of poly(3-hexylthiophene) (P3HT) and [6,6]-phenyl-C61-butyric acid methyl ester (PCBM). The structure of the POM layers depends on the thickness and shows a highly anisotropic surface organization. The films have been characterized by atomic force microscopy and X-ray photoelectron spectroscopy (XPS) to gain insight into their macroscopic organization and better understand their electronic properties. Then, they were put at the anodic interface of a P3HT:PCBM organic solar cell and characterized on an optical bench. The photovoltaic efficiency is discussed in terms of the benefit of the polyoxometalate at the anodic interface of an organic photovoltaic cell.

## Introduction

Low cost-fabrication, scalability, light weight and flexibility have driven the development of Organic PhotoVoltaics (OPV)[1] with Power Conversion Efficiency (PCE) values competitive[2] to those of Dye Sensitized Solar Cells. However, improving the performance of the donor-acceptor photoactive heterojunctions is still an intensive research area.[3–6] Much effort is also devoted to a better understanding of the role of interlayer materials. Interfacial layers (IFL) can be used to tune the band alignment, enhance the built-in electric field, improve the morphology of the organic film and lower interfacial charge recombination through favorable physical and electrical electrode/polymer contacts.[7–11] Many materials have been tested in a conventional device configuration either as electron collecting layer (ECL) at the cathode or hole collecting layer (HCL) at the anode. The beneficial effect of LiF as ECL or Poly(3,4-ethylenedioxythiophene) Polystyrene sulfonate (PEDOT:PSS) as anodic HCL is, for example, commonly recognized. However, PEDOT:PSS is also corrosive for the Indium tin oxide (ITO) anode, with detrimental effects on the long-term stability of the solar cell. To avoid these phenomena, inverted structured solar cells are now being investigated.[12–14] Recently, oxides like $V_2O_5$, $MoO_3$, $WO_3$ and NiO have been successfully introduced as HCL, while $TiO_x$ and ZnO have rather been placed at the cathode as ECL.[15,16] Oxides are attractive because of their low cost, visible light transparency, mechanical and electrical robustness, potentially high charge carrier mobility, and low environmental impact.[17]
Among molecular oxides, polyoxometalates (POMs) have outstanding structural diversity and tunable electronic properties.[18] However, their potential for solar cell applications has been explored almost exclusively in the liquid phase.[19–22] In the $[SiW_{12}O_{40}]^{4-}$ modified zinc oxide photoanode built on ITO, operating in a conventional electrochemical cell in the presence of the $I_3^-/I^-$ electrolyte, the POM is incorporated as an electron acceptor to limit charge recombination.[23] Two examples report on the implementation of $[PW_{12}O_{40}]^{3-}$ ($PW_{12}$) in solid-state devices for optoelectronics: as an electron injection layer in a Hybrid Organic Light Emitted Diode,[24] or as an ECL in a conventional ITO/PEDOT-PSS/P3HT:PCBM-61/$PW_{12}$/Al polymer solar cell.[25] In both cases, enhanced efficiency was attributed to the energy level alignment at the electrode.
One parameter seldom studied is the influence of the morphology at the interface on the electrical properties, although the possibility for easy nanostructuration of ZnO and the resulting minimization of surface defects has been discussed.[14,26,27] In view of using POMs in OPV, this issue is all the more essential since processing is still non-trivial, dominated by electrostatic layer-by-layer assemblies or polymer embedding.[28,29] Spontaneous adsorption of POMs on Ag, Au, glassy carbon or HOPG electrodes is well known[30–33] but has not been reported on ITO. On the other hand, drop casting of hybrid-POM solution on methylated and hydroxylated silicon surfaces has led to a wide variety of architectures imaged by scanning force microscopy.[34] This prompted us to investigate the spin coating growth of $K_6[P_2W_{18}O_{62}]$ (hereafter noted $K_6$-$P_2W_{18}$) on ITO. The dependence of the thickness on the structuration of the film will be discussed.
Surfaces were characterized by atomic force microscopy (AFM) and their electrical properties measured using the current sensing mode. Subsequently, the electronic structure of the highly structured layer has been determined by X-ray photoelectron spectroscopy (XPS) and Ultra-violet Photoemission spectroscopy (UPS). Finally, the $K_6$-$P_2W_{18}$ IFL was introduced in an heterojunction with poly(3-hexylthiophene) (P3HT) and [6,6]-phenyl-C61-butyric acid methyl ester (PCBM) of an organic photovoltaic device. Its opto-electrical properties have been characterized.

## Materials and methods.

$K_6[P_2W_{18}O_{62}]$ was prepared according to the accepted literature procedure[35]. Its molecular structure is depicted in the schematic of Fig. 1a.

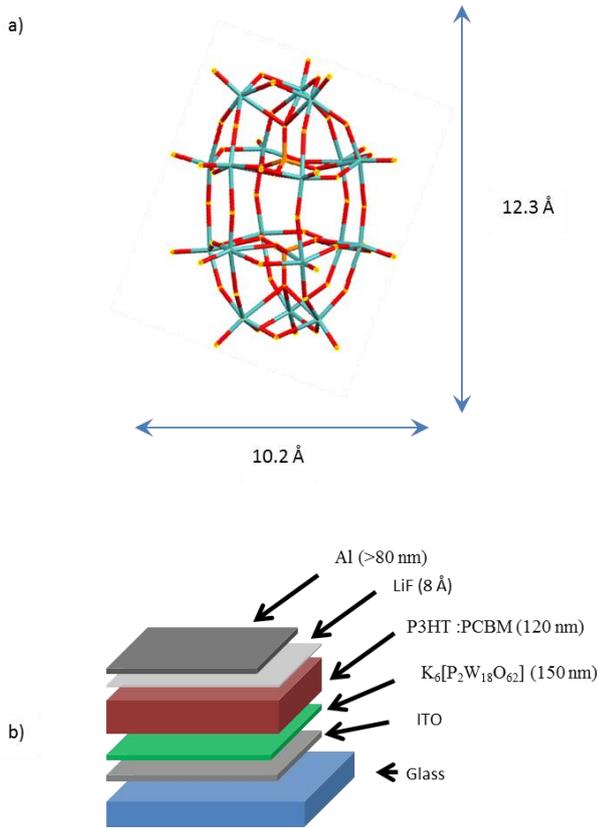

Figure 1: a) structure of the Wells- Dawson $K_6[P_2W_{18}O_{62}]$; b) Schematic of the photovoltaic cell

The POM films were grown onto a layer of indium tin oxide coated glass slide (75 Ω/sq, Sigma-Aldrich) in a glove box by spin coating in solution in dimethylsulfoxide (100 mg/mL) and annealed during 30 minutes at 140°C. The results for three layer thicknesses (40, 100 and 150 nm) are presented here. POM-based organic solar cells (OSC) were prepared using regioregular poly (3-hexylthiophene) (P3HT), [6,6]-Phenyl C61 butyric acid methyl ester (purity 99.5%, PCBM), poly(3,4-ethylenedioxythiophene)-poly(styrenesulfonate) (1.3 wt% in water, PEDOT:PSS), and lithium fluoride (assay 99.99%, LiF), all supplied by Sigma-Aldrich. The organic compound of OSC was deposited by spin coating, LiF and Al were deposited by thermal evaporation under high vacuum (P=$10^{-6}$ mbar) at rates of 0.01Å/s for LiF and 0.5Å/s for Al. With reference to the stack schematic in Fig. 1b the photovoltaic cells were prepared following the sequence: 1/ deposition of $K_6$-$P_2W_{18}$ (150 nm), 2/ spin coating of a mixture of P3HT:PCBM (respectively 15 mg/mL and 12 mg/mL in chlorobenzene), 3/ deposition of 7Å of LiF and finally 4/ deposition of the counter electrode of Al (100 nm). The process of layer deposition was performed under controlled nitrogen atmosphere. The layers were characterized by AFM on a Molecular Imaging from (Agilent, PicoLE), either in contact mode and current sensing mode (CS-AFM) with Pt/Ir tip (k=0.2N/m, radius = 20nm), the indentation force and surface contact were estimated at 20nN and 120 nm², respectively and the bias was applied to the ITO. The photo-electrical characterization of OSC was performed using a Xenon lamp, with a AMG1.5 filter calibrated at 75 mW/cm².
The XPS measurements were performed in an ultra-high vacuum system (base pressure $2 \times 10^{-10}$ mbar) using a monochromatic Al Kα X-ray source (1486.7 eV) and a SPHERA-Argus analyzer (both from Oxford Instruments Omicron Nanoscience). The overall energy resolution was better than 0.5 eV. The UPS measurements were made using an HIS-13 He I source (21.2 eV, also Oxford Instruments Omicron Nanoscience).

## Results and discussion.

**Morphological studies.**

The 40 nm POM film surface presented multiple spherulites with a lateral size between 500 nm and 1.5 μm (see Fig. 2a). Inspection of the AFM image shows the typical radial structure associated with spherulites. At this point, crystallization of the clusters has already started, however, the spherulites are still small and quite irregular with empty spaces between the domains. The line profile shown in Fig. 2b reveals grain heights up to 20 nm with an average value of 10 nm and the rms roughness is 5 nm. CS-AFM at an applied bias voltage of -500 mV showed a fully insulating behavior. At 100 nm, the size and the aspect of the spherulites became more regular and nearly constant (See Fig. 2c). The crystals are bigger (1μmx2μm) and have pentagonal or hexagonal forms, the growth is radial from a germination point (see profile line Fig. 2.d). The 3.9 nm rms roughness was smaller than that of thinner films and the grain boundaries were limited by the steric constraint. At the center of each spherulite a germination grain was clearly apparent from which the crystal grew radially (See Fig. 2c). At this stage of film growth, 2D steric effects reached a maximum. The surface was fully covered. The surface morphology of the 150 nm thick film changes dramatically. The spherulites have disappeared and the surface is fully covered by highly anisotropic ordered columns structured into domains, Fig. 2e. Within the domains a tubular organization was observed in two, orthogonal directions. The surface roughness was 3.7 nm. From the AFM line profiles shown in Fig. 2f, the column lengths reached a maximum of 3.4 μm, the maximum height was 6.6 nm and the width 100 (+/-10) nm. Line profiles give direct access to the surface topography, however, for a more quantitative measurement a 1D autocorrelation (formula 1) was performed. This analysis provided critical values for the height and the width of the columns (See Fig. 2g) which can be related to the size of $K_6$-$P_2W_{18}$. The resulting calculations gave $\Delta H_x$, the typical average height on the surface,

$$\Delta H_x = \sqrt{< \left(h(x + \Delta x) - h(x)\right)^2 >} \qquad (1)$$

where $h$ is the height at the $x$ position. The characteristic height was 2.1 nm which is the minimum height measured on the profile line (X1) in Fig. 2f. The width as estimated by the autocorrelation analysis was 48 nm, compatible with 32 elementary units assuming that the spacing between two units is around 3 Å[36] and that the POMs lie flat on the surface (32*(1.2+0.3)=48 nm). The cross section of the column read from the profile line is hence composed of 64 $K_6$-$P_2W_{18}$.

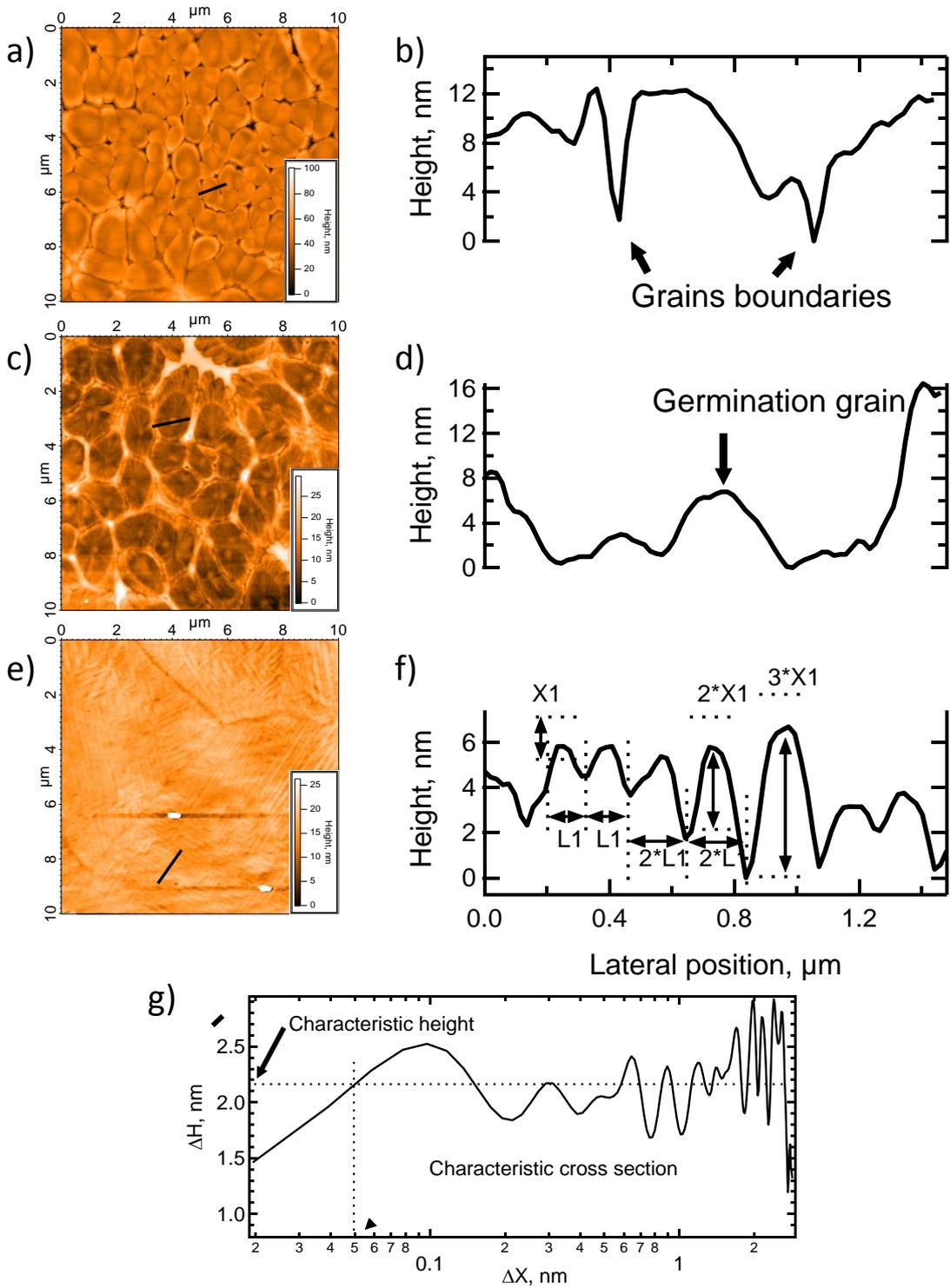

Figure 2: atomic force microscopic images of thin layers of $K_6[P_2W_{18}O_{62}]$; the black line inside the AFM images represents the location of the profile line on the right (a) for a thickness of 40nm the first steps of crystallization are observed, (b) profile of image (a) showing that the surface is rough, the grains are irregular and the grains boundaries are thick; (c), for a thickness of 100nm: the 2D growth crystal has reached a maximum limited by the steric constrain, the shapes of the crystal are regular and a germination grain at the center of the spherulite is present, (d) profile line of (c); (e) for 150nm film, the surface is composed of domains in which highly ordered columns oriented at 90° are present, (f) profile line of image (e) the columns clearly have characteristic parameters of wideness L1 an height X1. Those parameters were revealed by doing an auto-correlation analysis (g).

**Local electrical properties.**

The topography and the electrical behavior of the 150 nm film have been investigated by conductive AFM. At positive bias the surface was insulating. However, at -500 mV applied bias, and contrary to the 40 nm film, the surface became conductive, as shown in Fig. 3. Comparison of the topography with the electrical mapping indicates that the conductive pathways were mainly on top of the columns (Figure 3b and 3d). It is interesting to note the bright area seen at the center of the surface, Fig. 3c and 3d, was highly conductive. Thick layers of semiconductors (organic and/or inorganic), are known to have high resistivity (and/or poor conductivity). Only thin layers or highly crystalline structures show good electrical conductivity. Thus the conductivity of the 150 nm thick layer suggests that the bulk of the film is well-ordered and the bright area in the center of the images has a very high surface crystallinity.

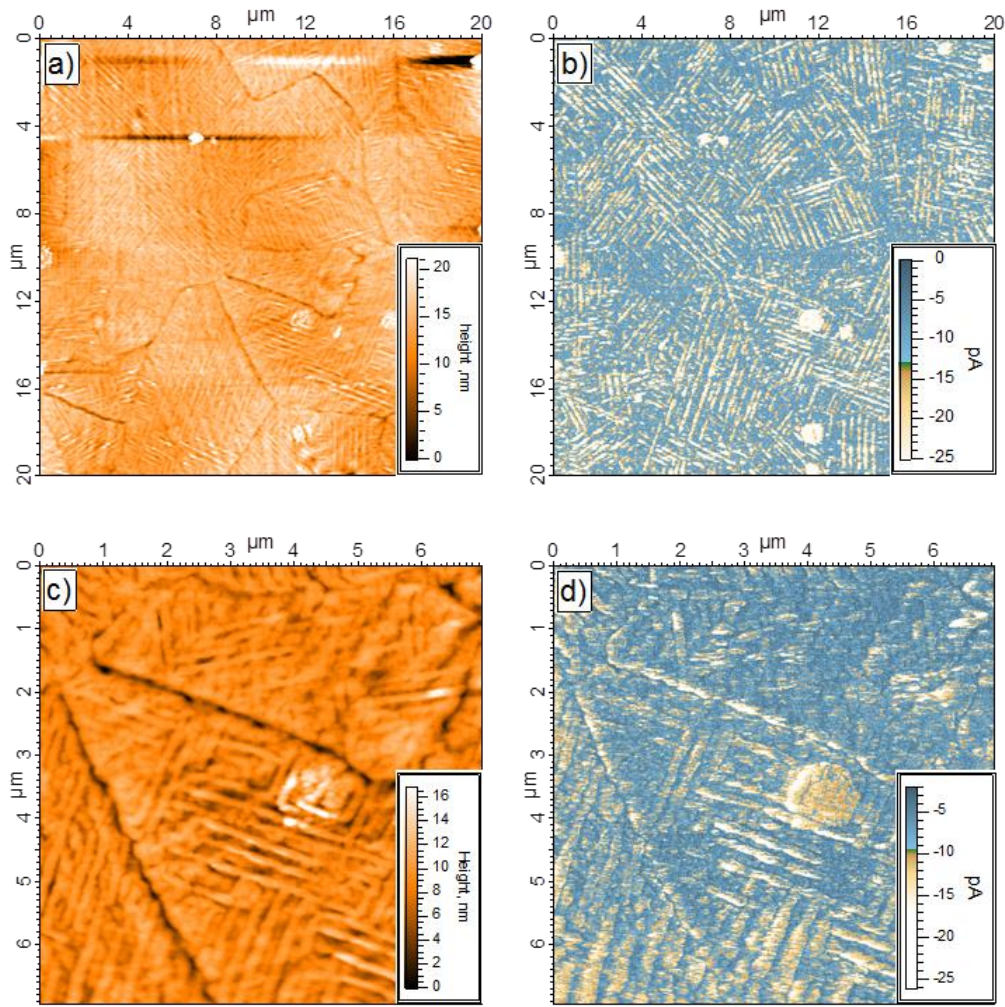

Figure 3: Topographic and current mapping at -500 mV applied bias on the 150 nm thick layer of $K_6[P_2W_{18}O_{62}]$; (a) Topography and (b) current mapping 20*20µm²; (c) Topography and (d) current mapping 7*7µm²

In addition to electrical mapping, local spectroscopy was performed to characterize the electronic structure of the layer (see Fig. 4b). The local I-V characteristics show a typical rectifying behavior. The current flow at negative bias shows a series of steps at -0.7 V and -1.0 V. At positive bias the I-V response is typical of a Schottky contact with a current flow beginning between 0.5 and 1.0 V with a current ratio at 1.5V calculated at 17. Due to empty d-levels, POMs are generally considered as electron acceptors. However, the current versus applied bias (I-V) measurements performed by CS-AFM demonstrated a high hole carrier mobility to ITO. Indeed, in the setup, at negative bias the electron injection was from ITO through the POM layer to the tip, thus ascribing a hole conducting behavior to POM layer. There might be a succession of discrete electronic states which progressively become more accessible as the magnitude of the negative bias is increased. On the contrary, there was low current at positive bias, the system blocked electron flow from the tip through the POM layer to ITO.

From the topographical and electrical mapping, it appears that the column-like structure in the domains reflects a well-ordered bulk structure as well as a highly-structured surface. Only ordering of the layer at the surface and in the bulk allows good electrical behavior (see Fig. ESI.1) Despite the thickness of the $K_6$-$P_2W_{18}$ layer, the surface was still conductive at low bias.

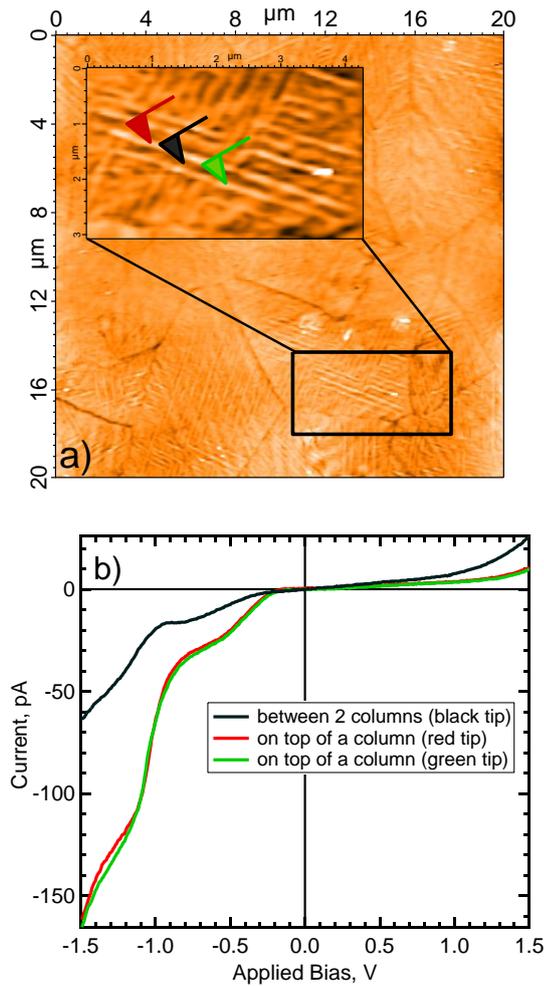

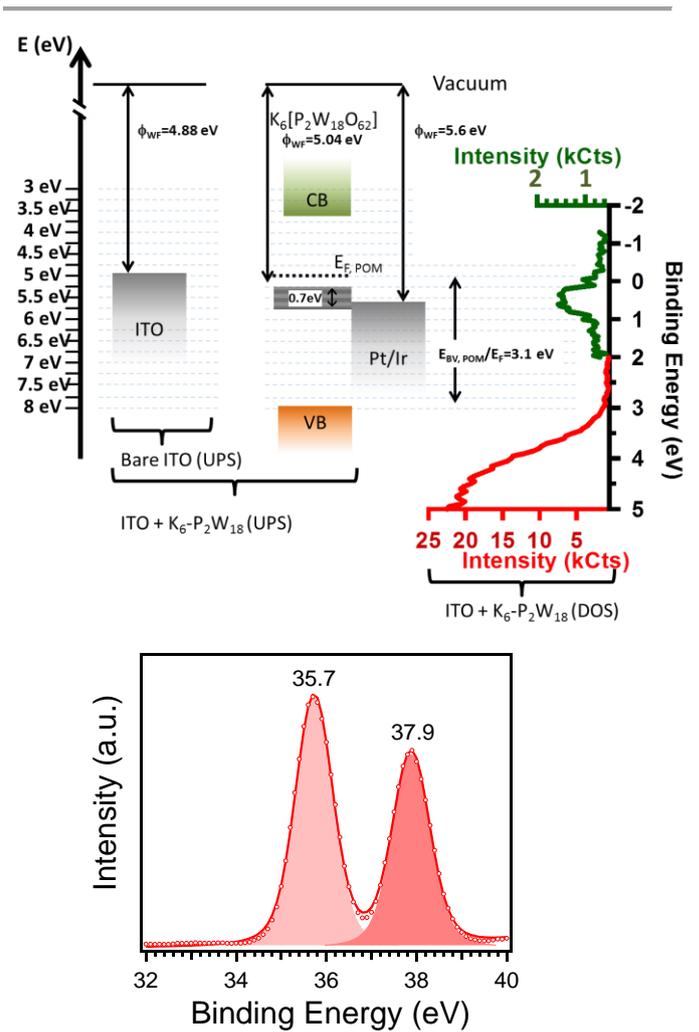

Figure 4: local I vs V spectroscopy using AFM (a) topographic surface with tips localization (b) spectroscopy

To better understand this behavior, X-ray and ultra-violet spectroscopy photoemission experiments were conducted to determine the electronic structure of the layer, the oxidation state of the POMs and the band alignment.

**Electronic structure**

The work function of a metallic sample surface can be directly obtained by measuring the threshold of the photoemission spectrum[37,38], i.e. the energy at which photoelectrons can escape from the material, measured with respect to the sample Fermi level. Figure ESI.2 shows the photoemission threshold of a thick, well-organized film of $K_6$-$P_2W_{18}$, as measured by UPS. Fitting the rising edge of the threshold by using an error function[39,40] gives a work function of 5.04 (0.020 eV) for the $K_6$-$P_2W_{18}$ layer. The work function of bare ITO has been measured at 4.88 eV (not shown).

Figure ESI.3 shows the valence band spectrum of a 150 nm thick layer of $K_6$-$P_2W_{18}$. The spectrum was acquired using XPS because the high photon energy reduces the contribution of secondary electron tail to the valence band emission allowing a clearer view of the valence band maximum (VBM) and localized states in the band gap. Figure 5a shows the energy band diagram for ITO and a $K_6$-$P_2W_{18}$ layer as deduced from the UPS measurements of the work functions and the XPS characterization of the valence band.

Figure 5: By using XPS and UPS we are able to obtain (a) the energy diagram of a 150 nm thick, well-structured layer of $K_6$-$P_2W_{18}$ onto ITO. The measurements gives the position of the maximum valence band at 3.1 eV below the Fermi level and show the presence of a broad gap state just below the Fermi level at 0.7 eV. The work function of AFM tip (Pt/Ir) was given by the literature[41]; red and green colors were used to evidence the change of intensity scale to highlight the gap state b) XPS W4f core-level spectrum showing a single component with $4f_{7/2}$ ($4f_{5/2}$) binding energy of 35.7 (37.9) eV

The VBM is 3.10 eV below the Fermi level of ITO. The signal visible just below the Fermi level might be due to photoelectrons emitted from the ITO substrate through the pores of the molecules, however, the film thickness and the absence of clear holes in the CS-AFM images suggests that the intensity is rather due to metallic like states localized in energy. The optical gap of the POM has been measured at 4.5 eV which allows us to locate the conduction band minimum at 3.64 eV (see Figure 5a), more typical of an n-type semiconductor.

The conducting behavior of the present POM layer is probably due to the presence of this intermediate energy level in the gap. Without the presence of these gap-states the POM layer would be fully resistive. This upholds the conductivity (current mapping) and the electronic response (local spectroscopy).

The question then is whether these in gap states are intrinsic to the POM or come from, for example, some adventitious reduction of the W, resulting in doping[16].

To get more insight into the electronic state of the POMs in the layer, we have measured core-level spectrum of the POM. The W $4f_{7/2}$ and $4f_{5/2}$ binding energies were 35.7 eV and 37.9 eV respectively, in agreement with previous values for $W^{6+}$ (see Fig. 5b). The core-level spectrum does not show any evidence for a second component which might be attributed to $W^{5+}$, excluding the possibility of significant reduction of the $W^{6+}$ [15] and adventitious doping.

Local spectroscopy suggests hole carrier conduction mechanism in the $K_6$-$P_2W_{18}$ layer, whereas the band alignment as measured by UPS and XPS is closer to an *n*-type electronic structure. This might suggest that the gap states are not fully populated allowing hole migration. This original behavior has been confirmed by the photovoltaic measurements.

**Photovoltaic response**

The photovoltaic response of OSC with an active layer of P3HT:PCBM and a 150 nm $K_6$-$P_2W_{18}$ IFL (see Fig. 1b) is presented in Fig. 6a. Figure 6b is a schematic of the expected band alignment in the stack, based on the known energy levels for the active layer and the energy diagram of Fig. 5. The photovoltaic cell has reached 2.6% of power conversion efficiency (PCE) using a solar simulator calibrated at 75 mW (Fig. 6a). The open circuit voltage and the fill factor are 440 mV and 42.5%, respectively. Moreover the I-V characteristic gives a very low series resistance of 15 Ω.cm² which contributes to the good fill factor value. It also suggests that $K_6$-$P_2W_{18}$ layer is conductive enough to drain charges from ITO to the active layer. The current density was 10.3 mA/cm². These results were compared to a reference cell using PEDOT:PSS as interfacial layer prepared in the same conditions and which showed a lower PCE at 1.5% (See Figure ESI.4).

The presence of discrete energy states in the gap, shown by both CS-AFM local spectroscopy and the UPS/XPS experiments, seems therefore to provide the channel for charge transport from the P3HT to ITO through the $K_6$-$P_2W_{18}$ layer.

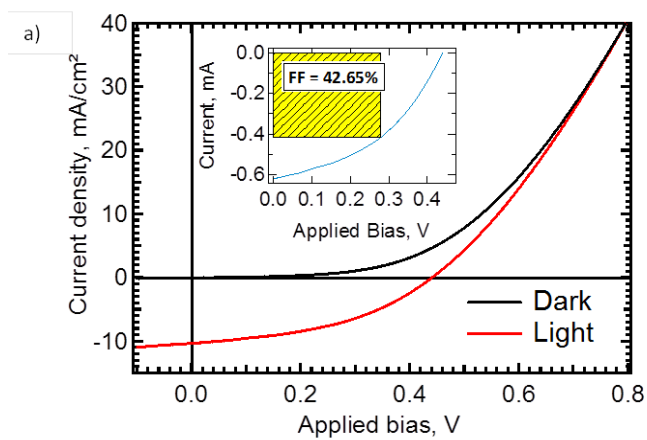

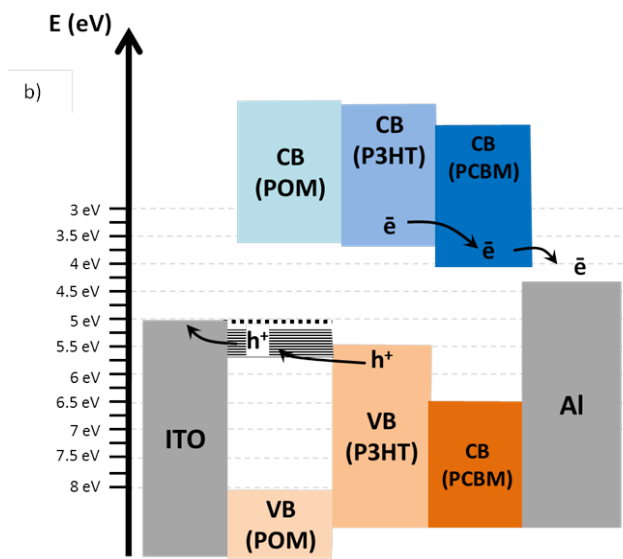

Figure 6: a) I vs V under 75 mW and b) full energy diagram

## Conclusions

Highly structured films of $K_6$-$P_2W_{18}$ on ITO were grown by spin-coating. The crystallization depends on the thermal annealing treatment and the film thickness. Well-ordered surface organization with anisotropic columns at 90° inside micrometric sized domains is obtained for films thicker than 150 nm. Electrical mapping reveals a conductive behavior for such layers whereas thinner layers are insulating. Local spectroscopy and UPS/XPS measurements suggest the presence of a discrete in-gap states which may facilitate charge transport. Both results point to a hole conducting behavior.

An organic P3HT:PCBM photovoltaic cell with a thick film of $K_6$-$P_2W_{18}$ at the anodic interface with ITO was compared to a reference with a PEDOT:PSS IFL. The cell shows a better efficiency (2.6 vs 1.5%), with excellent current density and good fill factor, and an optimized open circuit voltage. To gain more insight into the nature of the electronic states inside the band gap of the $K_6$-$P_2W_{18}$ layer, we are currently investigating other polyoxometalates of interest. Additionally, we will build working field effect transistor device to quantify the charge carrier selectivity and mobility.

## Acknowledgements

The authors would like to thanks the région Ile de France and the DIM Nano-K for funding the PhD grant to Qirong Zhu; J.E.R. was funded by CEA Ph.D. grants and by the Labex PALM APTCOM project

## Notes and references

*a* Sorbonne Universités, UPMC Univ Paris 06, CNRS UMR 8232, Institut Parisien de Chimie Moléculaire, F-75005 Paris, France. Fax:+33 1 44 27 38 41; Tel:+33144273034; E-mail : anna.proust@upmc.fr; e-mail : ludovic.tortech@upmc.fr

*b* CEA Saclay, IRAMIS, NIMBE, LICSEN, F-91191 Gif-sur-Yvette, France.

*c* CNRS, UMR 8232, Institut Parisien de Chimie Moléculaire, F-75005, Paris, France

*d* Synchrotron-SOLEIL, BP 48, Saint-Aubin, F91192, Gif sur Yvette CEDEX, France.

*e* CEA Saclay, IRAMIS, SPEC, LENSIS, F-91191 Gif-sur-Yvette, France.

†Electronic Supplementary Information (ESI) available: Local spectroscopy for different $K_6[P_2W_{18}O_{62}]$ layers, Spectrum of the threshold of photoemission, XPS binding energy in function of energy for a thick layer of $K_6[P_2W_{18}O_{62}]$ and I/V curve for photovoltaic reference cell. See DOI: 10.1039/b000000x/